\documentstyle[epsfig,axodraw]{aipproc}

\def\asymp#1%
{\mathrel{\raisebox{-.4em}{$\widetilde{\scriptstyle #1}$}}}

\def\Nequal#1%
{\mathrel{\raisebox{-.5em}{$\stackrel{=}{\scriptstyle\rm#1}$}}}
\newcommand{\dsl}[1]{\not \hspace{-0.7mm}#1}
\def\dsl{\mathpalette\make@slash}
\def\make@slash#1#2{\setbox\z@\hbox{$#1#2$}%
  \hbox to 0pt{\hss$#1/$\hss\kern-\wd0}\box0}

\newcommand{\rd}{{\mathrm{d}}}

\def\mathswitchr#1{\relax\ifmmode{\mathrm{#1}}\else$\mathrm{#1}$\fi}

\newcommand{\PW}{\mathswitchr W}

\newcommand{\Pe}{\mathswitchr e}

\newcommand{\Pep}{\mathswitchr {e^+}}
\newcommand{\Pem}{\mathswitchr {e^-}}

\def\mathswitch#1{\relax\ifmmode#1\else$#1$\fi}

\newcommand{\MW}{\mathswitch {M_\PW}}

\newcommand{\Me}{\mathswitch {m_\Pe}}

\def\solid{\raise.9mm\hbox{\protect\rule{1.1cm}{.2mm}}}
\def\dash{\raise.9mm\hbox{\protect\rule{2mm}{.2mm}}\hspace*{1mm}}

\def\ie{i.e.\ }

\newcommand{\sing}{{\mathrm{sing}}}
\newcommand{\finite}{{\mathrm{finite}}}

\newcommand{\DPA}{{\mathrm{DPA}}}

\newcommand{\Born}{{\mathrm{Born}}}
\newcommand{\IBA}{{\mathrm{IBA}}}

\newcommand{\virt}{{\mathrm{virt}}}
\newcommand{\real}{{\mathrm{real}}}

\newcommand{\eeWWffff}{\Pep\Pem\to\PW\PW\to 4f}
\newcommand{\eeffff}{\Pep\Pem\to 4f}
\newcommand{\eeffffg}{\eeffff\gamma}

\newcommand{\GeV}{\unskip\,\mathrm{GeV}}

\begin{document}

\renewcommand{\thefootnote}{\fnsymbol{footnote}}
\thispagestyle{empty}
\strut\hfill ER/40685/962 \\
\strut\hfill UR-1625 \\
\strut\hfill hep-ph/0101268
\vfill
\vspace*{-1cm}
\begin{center}
{\Large\bf Predictions for \boldmath{$\eeWWffff(\gamma)$} at \\ [.3em]
a future linear collider}
\footnote[2]{To appear in the {\it Proceedings of the 5th 
International Workshop on Linear Colliders} (LCWS 2000), 
Fermilab, Batavia, October 24-28, 2000.  \\
This work was supported in part by the U.S. Department of Energy,
under grant DE-FG02-91ER40685 and by the U.S. National Science Foundation, 
under grant PHY-9600155.}
\\[1.4cm]
{\large A. Denner$^1$, S. Dittmaier$^2$\footnote{Present address: DESY, D-22603 Hamburg, Germany}, M. Roth$^3$,
and D. Wackeroth$^4$} \\[1.5em]
\parbox{10cm}{{\small 
$^1$Paul-Scherrer-Institut, CH-5232 Villigen PSI, Switzerland} 
\\ [0.5em]
{\small $^2$Universit\"at Bielefeld, D-33615 Bielefeld, 
Germany}} \\ [0.5em]
{\small $^3$Universit\"at Leipzig, 
D-04109 Leipzig, Germany} \\ [0.5em]
{\small $^4$University of Rochester, 
Rochester, NY 14627, USA}
\end{center}
\vspace*{2.5cm}
{\large\bf Abstract}\\[.2cm]
We describe the salient features of precise predictions for the
processes $\eeWWffff(\gamma)$ as obtained with the Monte Carlo
generator {\sc RacoonWW}\footnote[3]{Program available 
from http://www.hep.psi.ch/racoonww/racoonww.html}, 
including the complete ${\cal O}(\alpha)$
electroweak radiative corrections in the double-pole
approximation. Numerical results for some distributions are given
at the typical linear-collider energy $\sqrt{s}=500\GeV$. 
Moreover, we study the impact of the
non-universal electroweak corrections by comparing with results of an
improved Born approximation.

\clearpage

\setcounter{page}{1}
\renewcommand{\thefootnote}{\arabic{footnote}}

\title{Predictions for \boldmath{$\eeWWffff(\gamma)$} at \\ [.3em]
a future linear collider}

\author{A. Denner$^*$, S. Dittmaier$^{\dagger}$, M. Roth$^{\ddagger}$,
and D. Wackeroth$^{\P}$}
\address{$^*$Paul Scherrer Institut, CH-5232 Villigen PSI, Switzerland \\
$^{\dagger}$Theoretische Physik, Universit\"at Bielefeld, D-33615 Bielefeld, 
Germany \\
$^{\ddagger}$Institut f\"ur Theoretische Physik, Universit\"at Leipzig, 
D-04109 Leipzig, Germany \\
$^{\P}$Department of Physics and Astronomy, University of Rochester, 
Rochester, NY 14627, USA}

\maketitle

\begin{abstract}
We describe the salient features of precise predictions for the
processes $\eeWWffff(\gamma)$ as obtained with the Monte Carlo
generator {\sc RacoonWW}, including the complete ${\cal O}(\alpha)$
electroweak radiative corrections in the double-pole
approximation. Numerical results for some distributions are given
at the typical linear-collider energy $\sqrt{s}=500\GeV$. 
Moreover, we study the impact of the
non-universal electroweak corrections by comparing with results of an
improved Born approximation.
\end{abstract}

\section*{Introduction}
\setcounter{footnote}{0}
The measurements of the W-boson mass and the gauge-boson 
self-interactions in W-pair production at a future $\Pep\Pem$ linear 
collider (LC) will provide further important precision tests of the
Electroweak Standard Model. To match the experimental
accuracy, which will exceed the per-cent level,
the predictions for the observed cross sections of $\eeWWffff(\gamma)$
have to reach an accuracy at the level of some $0.1\%$. 
Thus, the inclusion of higher-order effects in perturbation theory is
crucial. Moreover, at the LC for the first time a precise direct
measurement of quartic gauge-boson couplings 
can be performed, and predictions for all processes $\eeffffg$ are
needed. Already at LEP2 with only a few $WW\gamma$ events, first
direct bounds on the quartic gauge-boson 
\looseness -1
couplings have been obtained \cite{lep2quartic}.

Full results for the processes $\eeffff$ at the one-loop level are out
of sight at present.
Fortunately, for W-pair production at LEP2 and at not too high LC 
energies%
\footnote{Above 0.5--$1\,$TeV at least the leading electroweak
logarithms at the two-loop level should also be taken into account.}%
, it is
sufficient for the envisioned theoretical precision to take into
account only radiative corrections to those contributions that are
enhanced by two resonant W-boson
propagators.  Different versions of this
so-called double-pole approximation (DPA) for the
radiative corrections to off-shell W-pair production have been
described in the literature \cite{Be98,yfsww,ku99,paper}.
Two of them have been implemented in the state-of-the art Monte Carlo
generators {\sc YFSWW} \cite{yfsww} and {\sc RacoonWW} \cite{paper}.
The MC program {\sc RacoonWW} also provides tree-level predictions for all
processes $\eeffffg$ for massless fermions \cite{photon}.

Here we briefly describe the radiative corrections
to off-shell W-pair production in the DPA as implemented in {\sc
RacoonWW} and study their impact on some distributions 
at a typical LC energy $\sqrt{s}=500\,$GeV. 
Moreover, we discuss the effect of the non-universal electroweak
corrections by comparing the full calculation in the DPA with an
improved-Born approximation (IBA) \cite{radcor} as implemented in {\sc
RacoonWW}.  More results for LEP2 and LC energies are provided
in Refs.\cite{paper,lep2,lep2mcws} and \cite{radcor,nlc}, respectively.
Refs.\cite{paper,lep2mcws} also contain a discussion of the intrinsic
theoretical uncertainty of our version of the DPA as well as comparisons
to results of other authors.
\looseness -1

\section*{Precise Predictions for \boldmath{$\eeWWffff(\gamma)$}}

The ${\cal O}(\alpha)$-corrected cross section to off-shell W-pair production
in DPA as implemented in {\sc RacoonWW} can be written as
follows \cite{paper}:
\begin{equation}\label{eq:crosssection}
\rd \sigma_{\mathrm{WW}} =
\rd \sigma_{\Born}^{\eeffff}+
\rd \sigma_{\virt,\finite,\DPA}^{\eeWWffff}
+\rd\sigma_{\virt+\real,\sing}^{\eeffff}
+\rd\sigma_{\finite}^{\eeffffg}.
\end{equation}
Here $\rd\sigma_{\Born}^{\eeffff}$ is the full lowest-order cross
section to $\eeffff$, and $\rd\sigma^{\eeffffg}_{\finite}$, which describes 
real photon radiation away from IR and collinear regions, is the full
lowest-order cross section to $\eeffffg$ as described in Ref.\cite{photon}. 
The IR-finite sum of 
virtual and real soft and collinear photonic
corrections is denoted by $\rd\sigma_{\virt+\real,\sing}^{\eeffff}$.
The matching of IR and collinear singularities in the
virtual and real corrections is performed in two different ways: 
by using phase-space
slicing or alternatively a subtraction method \cite{subtraction}.  
Apart from non-mass-singular 
terms, $\rd\sigma_{\virt+\real,\sing}^{\eeffff}$
contains only collinear singularities associated with the initial
state, \ie leading logarithms of the form $\alpha\ln(s/\Me^2)$, at least for
sufficiently inclusive observables.  Since the contribution
$\rd\sigma_{\virt+\real,\sing}^{\eeffff}$ is not treated in DPA, those
logarithms are included in our approach exactly.  In {\sc RacoonWW},
the DPA is only applied to the finite (non-leading)
part of the virtual corrections,
denoted by $\rd\sigma_{\virt,\finite,\DPA}^{\eeWWffff}$ including the
full set of the remaining
factorizable and non-factorizable virtual ${\cal O}(\alpha)$ 
corrections. Beyond ${\cal O}(\alpha)$, {\sc RacoonWW} includes
soft-photon exponentiation and leading higher-order initial-state
radiation (ISR) up to ${\cal O}(\alpha^3)$ via the structure-function
approach (see Ref.\cite{lep2rep} and references therein). 
The leading higher-order effects from $\Delta
\rho$ and $\Delta \alpha$ are included
by using the $G_{\mu}$ scheme.  QCD corrections are taken into account
either by a multiplicative factor $(1+\alpha_{\mathrm s}/ \pi)$ for
each hadronically decaying W boson or by evaluating the 
${\cal O}(\alpha_{\mathrm{s}})$ corrections directly from 
\looseness -1
Feynman diagrams with real or virtual gluons.

\begin{figure}
{\centerline{
\setlength{\unitlength}{1cm}
\begin{picture}(14,6.8)
\put(-4.2,-15.){\includegraphics{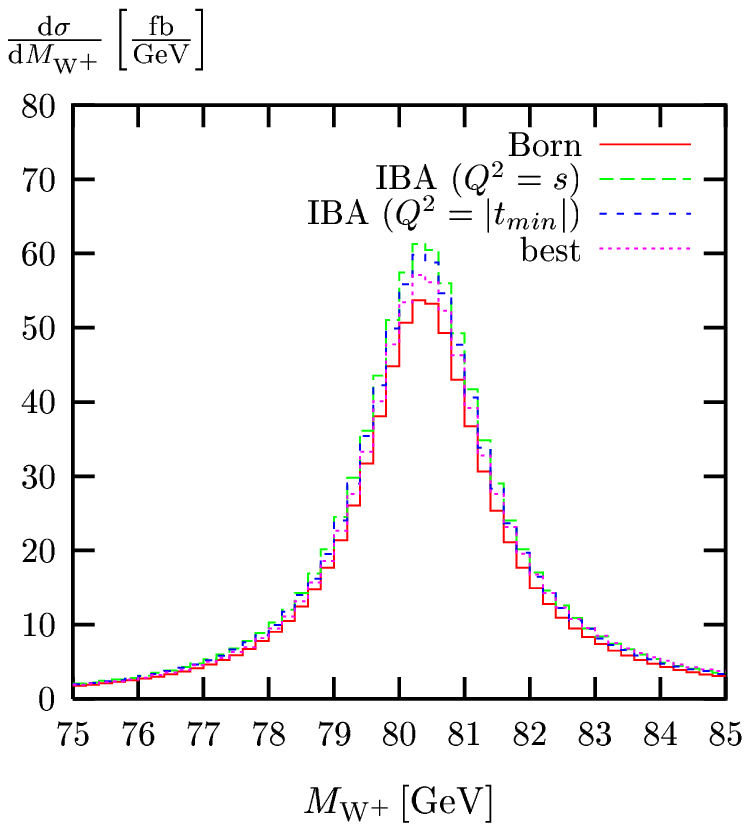}}
\put( 2.8,-15.){\includegraphics{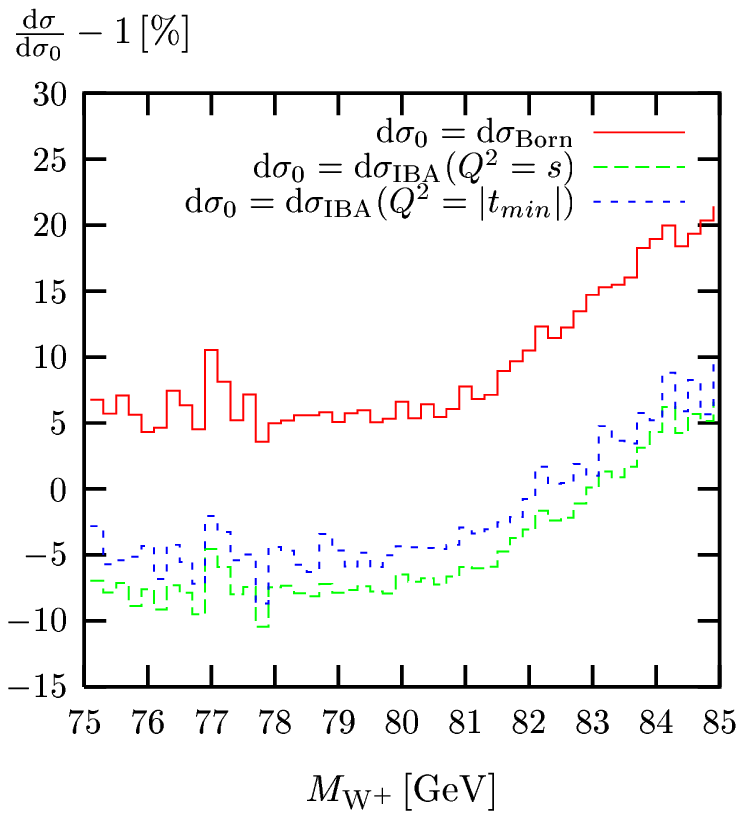}}
\end{picture} }}
\vspace{10pt}
\caption[]{The CC03 Born, IBA, and ``best'' predictions of the
$\mathrm{W}^+$ invariant-mass
distribution for ${\mathrm{e^+ e^- \to u \bar d}} \mu^-
\bar\nu_{\mu}$ at $\sqrt{s}=500\,$GeV when {\it calo} cuts are applied
(l.h.s). The corresponding relative corrections are shown for different
normalizations $\rd\sigma_0$ (r.h.s).}
\label{F:dwackeroth:1}
\vspace*{1em}
{\centerline{
\setlength{\unitlength}{1cm}
\begin{picture}(14,6.8)
\put(-4.2,-15.){\includegraphics{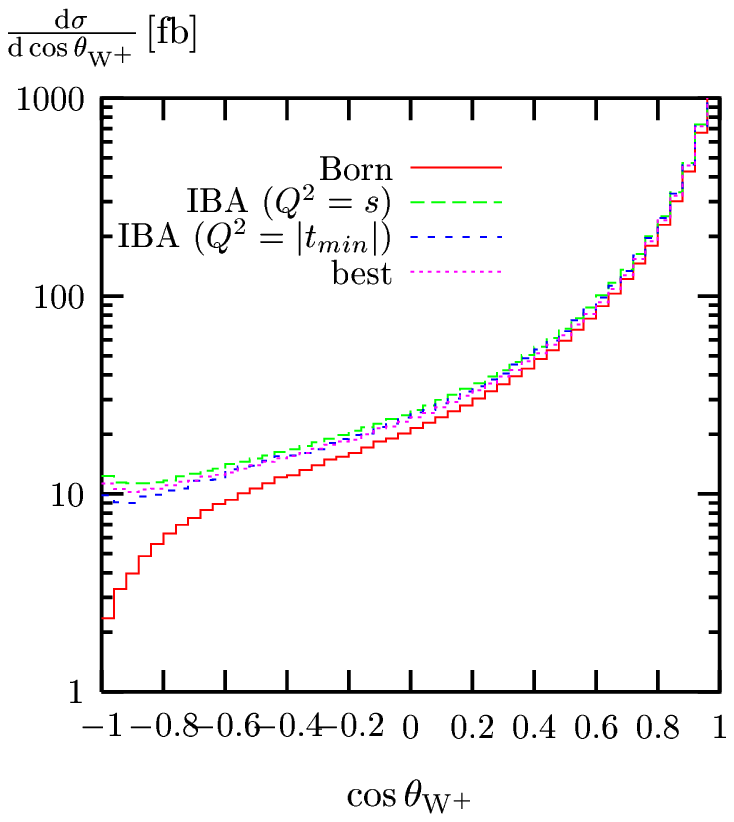}}
\put( 2.8,-15.){\includegraphics{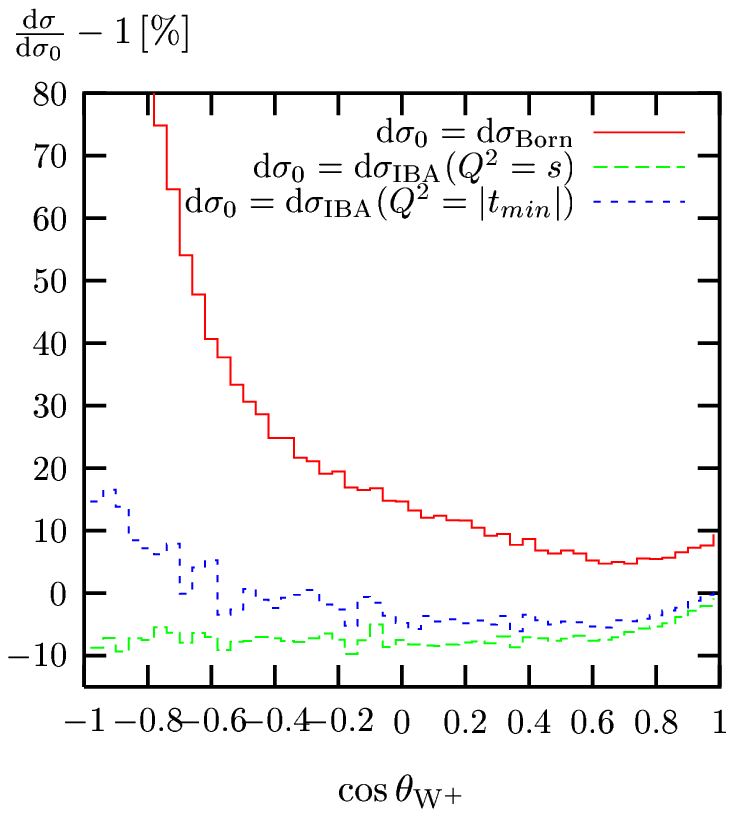}}
\end{picture} }}
\vspace{10pt}
\caption[]{The CC03 Born, IBA, and ``best'' predictions of the
distribution in the cosine of the $\mathrm{W}^+$ production angle
 at 500 GeV when {\it calo} cuts are applied (l.h.s).  The
corresponding relative corrections are shown for different
normalizations $\rd\sigma_0$ (r.h.s).}
\label{F:dwackeroth:2}
\end{figure}%
For the numerical results we use the setup of Ref.\cite{paper},
choosing the {\it calo} recombination procedure.
Our ``best'' results comprise the CC03 Born cross
section, $\rd\sigma_{\Born}$, and all radiative
corrections briefly described above.  We compare these
``best'' results with an IBA that has been implemented in
RacoonWW as described in detail in Ref.\cite{radcor}.  The IBA takes into
account the universal electroweak corrections \cite{bo92}: the running of the
electromagnetic coupling, corrections connected to the $\rho$
parameter, the Coulomb singularity, and leading-logarithmic ISR.

In Figs.~\ref{F:dwackeroth:1} and \ref{F:dwackeroth:2} we show the
$W^+$-invariant-mass and $W^+$-production-angle distributions,
respectively, and the corresponding relative corrections,
$\rd \sigma/ \rd\sigma_0-1$, to
${\mathrm{e^+ e^- \to u \bar d}} \mu^- \bar\nu_{\mu}(\gamma)$ at $500\GeV$. 
Here $\rd \sigma$ denotes the ``best'' prediction and 
$\rd\sigma_{\IBA}(Q^2)$ the IBA 
for two different scales in the structure functions, $Q^2=s$ and
$|t_{\mathrm{min}}|=s\left(1-\sqrt{1-4\MW^2/s}\right)/2-\MW^2$.  
In contrast to the LEP2 case, the radiative corrections
at $500\GeV$ mostly increase the Born cross sections.

The invariant masses are obtained from the four-momenta of the decay
fermions of the W bosons after eventual photon recombination.  As
discussed in Refs.\cite{paper,lep2}, they are very sensitive to the
treatment of the photons, i.e.\ different results are obtained when
using {\it bare} cuts.  The observed distortion of the
W-invariant-mass distributions is of particular interest when the
W-boson mass is reconstructed from the W-decay products.  As can be
seen when comparing the ``best'' with IBA results, this distortion
cannot be described by the used IBA, since the IBA does not account
for radiation from the W-decay processes.  The universal corrections
mainly affect the normalization of the invariant-mass distributions.

For the angular distributions, we define all angles in the 
centre-of-mass system of the initial state.  In contrast to
the invariant-mass distributions, the angular distributions hardly
depend on the recombination procedure. Thus, similar results are obtained
when {\it bare} recombination cuts are used \cite{paper,lep2}. The dramatic
increase of the relative corrections is due to hard ISR
which causes a redistribution of events to a
phase-space region where the Born cross section $\rd\sigma_{\Born}$ is
very small.  Thus, the effect is less pronounced when
$\rd\sigma_0=\rd\sigma_{\IBA}$, since the IBA accounts for the leading
ISR effects. For very small angles, where the $t$-channel pole dominates
the cross section, the IBA predictions approach the ``best'' results 
within a few per cent.
For large and intermediate angles, where the cross section is small,
the IBA predictions depend strongly on the scale in the structure
functions.
\vspace*{-1ex}\hfill

\end{document}